\documentclass[supp]{new_tlp} 

\usepackage{supp}
\usepackage{makeidx}  
\usepackage{graphicx}
\usepackage{caption}
\usepackage{subcaption}
\usepackage{listings}
\usepackage{float}
\usepackage{url}
\usepackage{footnote}
\usepackage{comment}
\restylefloat{figure}

\usepackage{longtable}

\usepackage[T1]{fontenc}
\usepackage[scaled]{beramono}
\newcommand\Small{\fontsize{8}{8.2}\selectfont}

\newcommand\bcmdtab{\noindent\bgroup\tabcolsep=0pt%
  \begin{tabular}{@{}p{10pc}@{}p{20pc}@{}}}
\newcommand\ecmdtab{\end{tabular}\egroup}

\newcommand{\chr}{\textsf{CHR}}

\begin{document}
\bibliographystyle{acmtrans}

\long\def\comment#1{}

\title{Visualization of Constraint Handling Rules}
 
\author[Nada Sharaf, Slim Abdennadher and Thom Fr{\"u}hwirth]{Nada Sharaf\textsuperscript{1}, Slim Abdennadher\textsuperscript{1} and Thom Fr{\"u}hwirth\textsuperscript{2}
\\\textsuperscript{1}The German University in Cairo, Egypt; \textsuperscript{2}Ulm University, Germany
\\(e-mail:\{nada.hamed, slim.abdennadher\}@guc.edu.eg, thom.fruehwirth@uni-ulm.de)
}

\pagerange{\pageref{firstpage}--\pageref{lastpage}}
\volume{\textbf{10} (3)}
\jdate{March 2002}
\setcounter{page}{1}
\pubyear{2002}

\pubauthor{Sharaf,Abdennadher,Fruehwirth}
\jurl{xxxxxx}
\pubdate{22 June 2013}\maketitle

\label{firstpage}

 \begin{abstract}
  
    Constraint Handling Rules (\textsf{CHR}) has matured into a general purpose language over the past two decades. Any general purpose language requires its own development tools. Visualization tools, in particular, facilitate many tasks for programmers as well as beginners to the language. The article presents on-going work towards the visualization of \textsf{CHR} programs. The process is done through source-to-source transformation. It aims towards reaching a generic transformer to visualize different algorithms implemented in \textsf{CHR}.
\\Note: An extended abstract / full version of a paper accepted to be presented at the Doctoral Consortium of the 30th International Conference on Logic Programming (ICLP 2014), July 19-22, Vienna, Austria.
    
  \end{abstract}

\section{Introduction}

Although Constraint Handling Rules (\textsf{CHR})\cite{Fru98} was introduced as a language for writing constraint solvers, it has developed into a general purpose language over the years. 
CHR is a committed choice language. A \chr\ program consists of multi-headed guarded rules. In \textsf{CHR}, predicates are transformed into simpler ones until they are solved. \textsf{CHR} has a number of implementations. However, the most prominent ones are embedded in Prolog.

Since \chr\ has developed into a general purpose language, \chr\ programmers can now write programs to implement general algorithms such as sorting algorithms, graph algorithms, etc. With such algorithms, programs could get very long and in some cases complicated. Thus development tools and especially tracing and visualization tools are very useful and sometimes even necessary.
As discussed in \cite{DBLP:journals/vlc/HundhausenDS02}, algorithm visualization technologies are useful in many cases such as in practical laboratories, for in-class discussions, or in assignments where students could for example produce their own visualizations. It could help instructors find bugs quickly. Moreover, such visualizations could be useful for
debugging and tracing the implementations of different algorithms.

There are different methods for embedding visualization features into a \chr\ program. One of them is to alter the compiler or the \chr\ runtime system. This solution is however not recommended since performing such changes is not an easy task for any programmer especially a beginner. Thus the adopted approach is to use source-to-source transformation to eliminate the need of doing any changes to the running system.

Program transformation or source-to-source transformation, allows developers to add or change the behavior of programs without manually
modifying the initial code. In addition, according to \cite{Loveman:1977:PIS:321992.322000}, source-to-source transformation
could be useful for improving the performance of different programs

In \cite{stschrpaper}, a first attempt towards visualizing \chr\ programs was presented. The presented tool was able to visualize the execution of \chr\ programs was realized through source-to-source transformation. In addition, it was able to visualize \chr\ constraints as objects. The system, however, lacked generality and required ad-hoc hard-wired inputs. In addition visualizing algorithms implemented through the different \chr\ programs was not possible.
Some systems \cite{oz,tree} provided visualization options for constraint programs. However, the focus was on the search space and trees rather than the executed algorithms.
This paper thus presents a new approach that aims at visualizing \chr\ solvers in a generic way overcoming the problems of the old tool. Through the new system, various algorithms implemented through \chr\ could be visualized.

The final goal of the project is to have a general source-to-source transformation tool that is able to automatically add different extensions to \chr\ solvers. 
\\The paper is organized as follows: Section \ref{sec:chrintro} introduces \chr\ through an example. Section \ref{sec:archi} discusses in more details the suggested architecture of the system. An example of the output of the system is shown in Section \ref{sec:jawaasec}. Finally, some conclusions and directions for future work are shown in Section \ref{sec:conc}.

\section{Constraint Handling Rules}
\label{sec:chrintro}
This section presents an example of a \chr\ program to introduce the syntax and semantics of \textsf{CHR} \cite{chrbook}.
In \textsf{CHR}, two types of constraints are
available. The first type is the built-in constraints that are provided through
the host language. The second type of constraints is the \chr\ or user-defined
constraints that are defined through the rules of the program
A \chr\ program consists of \emph{simpagation rules} with the form:
\[name\;@\;H_{k}\;\backslash\; H_{r}\; \Leftrightarrow \; G \; |\; B. \]
The name of the rule precedes the \verb+@+ sign and is optional. The head of the \chr\ rule, comes before the (\(\Leftrightarrow\)). It should only contain a conjunction of \chr\ constraints.
As seen from the previous rule, there are two parts in the head namely \(H_{k}\) and \(H_{r}\).
\(H_{k}\) contains the constraints that are kept after the rule is executed.
However, the constraints in \(H_{r}\) are removed after executing the rule.
The guard (\(G\)) is optional and should only contain built-in constraints.
Finally, the body (\(B\)) could contain both \chr\ and built-in constraints. The constraints in the body are added to the constraint store on executing the rule.
\\A rule is executed only if the head constraints match some of the constraints in the constraint store \cite{fru_welcome_lnai08}. In addition, the guard has to be satisfied for the rule to be executed. At the beginning of the execution, the constraint store is empty. It is initialized by the constraints in the query.
\\There are special cases of simpagation rules which are simplification and propagation rules.
Whenever \(H_{k}\) is empty, the resulting rule is a  ``simplification'' rule. The head of a simplification rule contains \chr\ constraints that are removed once the rule is executed. Thus through \emph{simplification} rules, \chr\ constraints are replaced by simpler ones.
Thus the format of a simplification rule is:
\[ H_{r} \;  \; \Leftrightarrow \;  \; G  \;  \;  \; | \;  \;  \; B. \]
In \emph{propagation rules}, \(H_{r}\) is empty. Consequently, the rule does not remove any constraints from the store. It only adds the constraints in the body to the store. Propagation rules have the following format:
\[ H_{k} \;  \; \Rightarrow \;  \; G  \;  \;  \; | \;  \;  \; B. \]

The following example is for a \chr\ program that sorts numbers. The numbers are fed into the solver using the constraint \verb+list+. For example, the constraint \verb+list(I,V)+ means that the cell at index \verb+I+ of the list has the value \verb+V+. The solver contains 
the simplification rule \verb+sortlist+.
\begin{verbatim}
sortlist @ list(Index1,V1), list(Index2,V2) <=> Index1<Index2 , V1>V2 | 
                       list(Index2,V1), list(Index1,V2).
		\end{verbatim}
The rule \verb+sortlist+ makes sure that a number precedes another one in the list if and only if it is indeed smaller than it. If this is not the case, the two numbers are swapped. 
Consequently, applying such a rule results in a sorted sequence of numbers.
For example if the input to the solver is \verb+list(0,7), list(1,6), list(2,4)+, execution proceeds as follows:
\begin{center}
\(\underline{list(0,7), list(1,6)}, list(2,4)\)\newline
\(\Downarrow\)\newline
\(list(1,7), \underline{list(0,6), list(2,4)}\)\newline
\(\Downarrow\)\newline
\(\underline{list(1,7), list(2,6)}, list(0,4)\)\newline
\(\Downarrow\)\newline
\(list(2,7), list(1,6), list(0,4)\)\newline
\end{center}
As seen from the previous execution sample, every time a new number is added to the sequence or the store, the program makes sure it is placed in the correct position with respect to the already existing elements. Accordingly, after all numbers are added, the resulting sequence is a sorted one.
The semantics implemented in SWI Prolog is the refined operational semantics \cite{refined}. It makes sure that constraints are processed from the left to the right and that rules are executed in a top-bottom approach as demonstrated through the previous example.
For example after \verb+list(2,4)+ is added to the constraint store, the rule \verb+sortlist+ is executed since \verb+6+ and \verb+4+ are not sorted in the sequence.
Thus, \verb+list(0,6)+ and \verb+list(2,4)+ are removed from the constraint store. They should be replaced by \verb+list(2,6)+ and \verb+list(0,4)+. However, these two constraints are added to the store one by one. Once \verb+list(2,6)+ is added to the store and even before adding \verb+list(0,4)+, the rule \verb+sortlist+ is executed using the two constraints \verb+list(2,6)+, \verb+list(1,7)+.
A more detailed view of the execution steps is:
\\\\\\
\begin{center}
\begin{longtable}{cp{8cm}}
 An Empty Constraint Store & \(\) \\
 \(\Downarrow\) & Adding \verb-list(0,7)- to the store\\
 \(list(0,7)\) & \(\) \\
 \(\Downarrow\) & Adding \verb-list(1,6)- to the store\\
 \(\underline{ list(1,6),list(0,7)  }\) & \(\) \\
 \(\Downarrow\) & \verb-sortlist- removing \verb-list(0,7)-, \verb-list(1,6)- and adding adding \verb-list(1,7)- as a first step\\
  \(list(1,7)\) & \(\) \\
  \(\Downarrow\) & Adding \verb-list(0,6)- to the store\\
   \(list(0,6), list(1,7)\) & \(\) \\
   \(\Downarrow\) & Adding \verb-list(2,4)- to the store\\
    \(\underline{list(2,4), list(0,6)}, list(1,7)\) & \(\) \\
     \(\Downarrow\) & \verb-sortlist- removing \verb-list(2,4), list(0,6)- and adding adding \verb-list(2,6)- as a first step\\
     \(\underline{list(2,6), list(1,7)}\) & \(\) \\
      \(\Downarrow\) & \verb-sortlist- removing \verb-list(2,6), list(1,7)- and adding adding \verb-list(2,7)- as a first step\\
       \(list(2,7)\) & \(\) \\
        \(\Downarrow\) & Adding \verb-list(1,6)- to the store\\
        \(list(1,6),list(2,7)\) & \(\) \\
         \(\Downarrow\) & Adding \verb-list(0,4)- to the store\\
         \(list(0,4),list(1,6),list(2,7)\) & \(\) \\
\end{longtable}
\end{center}

\section{System Architecture}
\label{sec:archi}
\begin{figure}
\caption{The General Architecture}
\centering
\includegraphics[width=100mm]{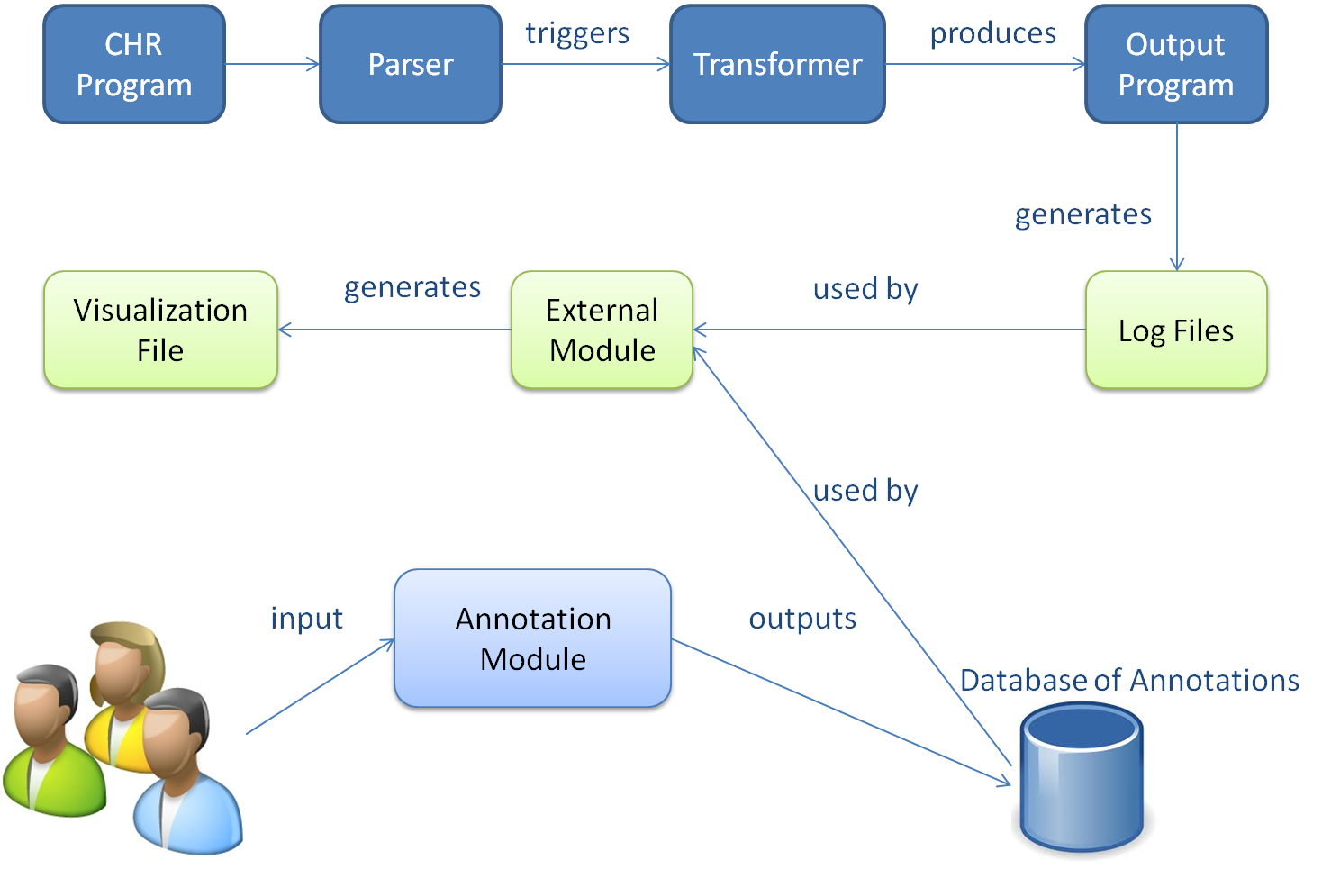}
\label{fig:general}
\end{figure}

This section introduces the adopted architecture and transformation approach. The tool presented by \cite{stschrpaper} was able to add visualization features to CHR programs. 
The tool however lacked the possibility of visualizing different algorithms implemented through \textsf{CHR}.
In order to extend the tool, the user was always required to enter specific hard-wired inputs.
%
The focus is now for a new and a more general approach. As shown in Figure \ref{fig:general}, the general architecture of the workbench consists of several modules. At the beginning, the \chr\ program is fed into the parser. In addition to parsing the input file, the parser also extracts the needed information and represents it in the required format for the transformer. 


The transformer uses the ``relational normal form'' presented in \cite{stsforexpresive}. This form uses some special \chr\ constraints to encode the different constituents of a \chr\ rule. Such constraints include \verb-head-, \verb-body- and \verb+guard+. The parser thus represents the information about the different rules using the specified form. For example, \verb+head(sortlist,`list(Index1,V1)',remove)+ represents the fact that the rule \verb+sortlist+ has the constraint \verb+list(Index1,V1)+ in its head. In addition, this constraint is removed on executing the rule since it is a simplification rule.
Finally, the new solver is generated by the transformer. Unlike the form presented in \cite{stsforexpresive}, the transformer neglects the identifier of the head constraint since it is not needed.

In addition, the system makes use of a new module called the ``Annotation Module'' explained in more detail in Section \ref{sec:annotation}.

The output programs are normal \chr\ programs that could run with SWI-Prolog. Generally, when the output solver is running, log files are generated as shown in Figure \ref{fig:general}. These files should contain information regarding the executed rules that could then be used by an optional external module.

The external module is utilized by the visualization extension. This module reads the log files generated by the new program. In addition it makes use of the output of the \emph{Annotation Module} in order to produce the visualization file needed by the visual tracer.
This visual tracer could be any visualization program. For proof of concept we used Jawaa \cite{jawaa}.
\subsection{Annotation Module}
\label{sec:annotation}
\begin{figure}[!ht]
  \centering
 \includegraphics[width=90mm]{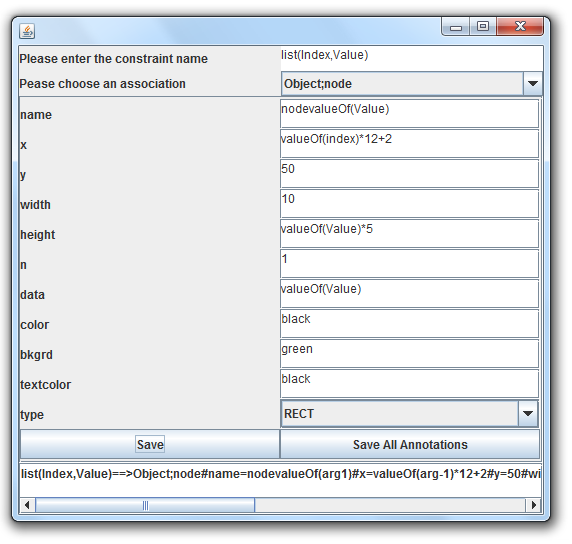}                
  \caption{Annotating CHR constraints.}
  \label{fig:jawaamapping}
\end{figure}

This module was proposed as a step towards eliminating the need of entering algorithm-specific information. The new approach this paper introduces also aimed at a more general visual tracer that does not need to be changed every time the algorithm differs. 
The idea is to have a more basic visualization tool and a more intelligent transformation process. Thus, we decided to out-source the actual visualization process to existing tools such as Jawaa, OpenSCAD\footnote{\url{http://www.openscad.org/}}, etc.
Such systems provide different sets of visualization objects and possibly actions as well.
Nevertheless, the need of connecting the transformed \chr\ programs to these systems in a generic way to be able to visualize any algorithm remained.
\\As introduced in \cite{bookintro}, algorithm animation or software visualization produces abstractions for the data and the operations of an algorithm. The different states of the algorithm are represented as images that are animated according to the different interactions between such states.
\\The adapted idea in the new system is to visualize the \chr\ constraints themselves as objects. After transforming the initial program, the rules of the new program modify the constraints and thus the objects. 
Consequently, the execution of each rule adds one step to the visual tracer. Viewing the sequence of objects thus produces an animation showing how the rules acted on the constraints (or objects) and thus visualizes the algorithm.
\\Accordingly, the ``\emph{Annotation Module}'' was suggested. The module extends the system with the annotation functionality which allows it to deal with different visualization tools while keeping a general scheme.
When deploying the system, users enter the annotations or the mappings between the different constraints and the objects provided through the visualization system. Figure \ref{fig:jawaamapping} shows how the user mapped the constraint \verb+list/2+ to the visual object ``Node'' that Jawaa offers. In addition to stating the name of the object, the user also states how the arguments of the constraints affect the parameters needed for the object. The parameters for the \emph{Node} object include the name, the x-coordinate, the y-coordinate, the width, the height in addition to the text. The values the user enters for these parameters are also associated with the annotations.
The values could use some of the arguments of the constraints.
As shown in Figure \ref{fig:jawaamapping}, the constraint is mapped into a \emph{Node} object which means that everytime a \verb+list/2+ constraint is added to the store a new \emph{Node} is visualized. The name of the \emph{Node} is chosen to be \emph{nodevalueOf(Value)} which means that the each element in the list will have a node with a name corresponding to its value. For example, the Node corresponding to the number \verb+7+ is a \verb+node7+.
The x-coordinate is calculated with the following formula \(valueOf(Index)*12+2\). Through this formula the Node of the element at index \verb-0- is placed at the x-coordinate \verb-2-. The Node of the second element (with index \verb-1-) is placed at the x-coordinate \verb-14-. The third element is placed at \verb+26+. Given that the width is determined to be 10, this means that there is a 2 pixels gap between any two consecutive nodes. The y-coordinate is determined to be \verb+50+ for all of the Nodes. The height, on the other hand, is five times the value of the element represented through \verb-valueOf(Value)*5-. The node contains only one piece of information which is the value of the element.
The outline color of the node is black and the background color is green.

Whenever the user saves the annotation, the following XML file is produced keeping track of all of the associations to the constraint.
  \lstset{basicstyle=\Small,
          breaklines=true
          }
\begin{lstlisting}[language=xml]
<association>
<constraint name="list(Index,Value)">
<add name="node" parameters="name=nodevalueOf(arg1)#x=valueOf(arg0)*12+2#y=50#
width=10#height=valueOf(arg1)*5#n=1#data=valuef(Value)#color=black#bkgrd=green#textcolor=black#type=RECT" type="arg1"/>
</constraint>
</association>
\end{lstlisting}
\subsection{Transformation Module}
This section shows how the rules in the original \chr\ program are transformed to be able to interact with the visualization system. Source-to-source transformation was used in order to avoid any manual changes.
The basic scheme is similar to the one shown in \cite{stschrpaper}.
In more details, the new solver interacts with an external module. This module was implemented in Java. The external module uses the information propagated through the solver in addition to the saved annotations to produce a file that could be animated through the visualization system such as Jawaa or OpenSCAD.
To have a step-by-step animation, the execution of each rule communicates the needed information.
\\In general any \chr\ rule of the form:
\[rule\; @\; H_{k}\; \backslash \; H_{r}\; <=>\; G\; |\; Body.\]
is transformed to a rule of the form:
\[rule\; @\; H_{k} \; \backslash \; H_{r}\; <=>\; G\; |\; communicate\_hk(H_{k}),\; communicate\_hr(H_{r}),\; Body.\]
As seen from the scheme, the functionality of the rule is kept intact. However, the auxiliary predicate \verb+communicate/1+ is used within the new body of the rule. The aim of this predicate is to send the head of the rule to the external module. This module can then start to act upon the information to add the next visualization step. 
\\Actually, only the heads that are removed from the constraint store can affect the visualization. In such a case, the visualized objects for these constraints should be removed from the visualization window.
\\In some cases, the constraints of the head will not affect the visualization. The transformer can then be instructed to produce a new solver that does not communicate to the external module the head constraints. In such output solvers the old rules are kept intact:
\[rule\; @\; H_{k}\; \backslash \; H_{r}\; <=>\; G\; |\; Body.\]
However, such transformation is not sufficient since the constraints in the body of the rule were not communicated. The body-constraints are responsible for adding new constraints/objects to the trace.
Since with the proposed system each constraint maps to an object, a generic way of solving this problem is to add for each constraint \verb+cons/n+ a rule with the form:
\[cons(arg_{1},\;arg_{2},\;arg_{3},\;\ldots\;,arg_{n})\Rightarrow communicate(cons(arg_{1},\;arg_{2},\;arg_{3},\;\ldots\;,arg_{n})).\]
The previous rule is a propagation rule which does not affect the constraint store. It is however triggered once a constraint of the form \(cons(arg_{1},\;arg_{2},\;arg_{3},\;\ldots\;,arg_{n})\) is added to the store communicating to the external module the new constraint in the store. 
If this constraint has a corresponding annotation, the module adds the needed visualization step.
In addition, if a rule adds multiple constraints that have corresponding annotations, they will be animated one by one. Thus the visualization step is done once the constraint is added to the store.
Such propagation rules are added at the beginning of the program to make sure they are executed once a constraint is added and before executing any other applicable rule.

\section{Jawaa Example}
\label{sec:jawaasec}
This section shows how the sorting algorithm shown in Section \ref{sec:chrintro} is visualized using Jawaa. 
The output solver has the following extra rule at its beginning:
\begin{verbatim}
list(V0,V1) ==> communicate(list(V0,V1)).
\end{verbatim}
In addition the rule \verb+sortlist+ is modified such that it communicates to the external module its head constraints since their corresponding objects need to be removed.
Thus the new rule has the following format:
\begin{verbatim}
sortlist @ list(Index1,V1), list(Index2,V2) <=> Index1<Index2 , V1>V2 | 
        communicate_hr(list(Index1,V1)), communicate_hr(list(Index2,V2)), 
        list(Index2,V1), list(Index1,V2).
\end{verbatim}
To have an animation through Jawaa, a ``anim'' file that contains all of the animation details needs to be used.
The external module makes use of the information communicated from the new solver in addition to the saved annotations in order to generate the corresponding animation file.
The external module is able to dynamically build up the animation file step by step.

\begin{figure}[!ht]
  \centering             
\begin{subfigure}[b]{20mm}\includegraphics[width=20mm]{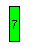}\caption{adding\newline list(0,7)}\end{subfigure}%
  \begin{subfigure}[b]{20mm}\includegraphics[width=20mm]{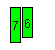}\caption{adding\newline list(1,6)}\end{subfigure}%
  \begin{subfigure}[b]{20mm}\includegraphics[width=20mm]{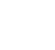}\caption{removing\newline list(0,7) ,\newline list(1,6)}\end{subfigure}%
   \begin{subfigure}[b]{20mm}\includegraphics[width=20mm]{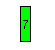}\caption{adding\newline list(1,7)}\end{subfigure}%
   \newline\begin{subfigure}[b]{20mm}\includegraphics[width=20mm]{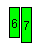}\caption{adding\newline list(0,6)}\end{subfigure}%
  \begin{subfigure}[b]{20mm}\includegraphics[width=20mm]{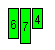}\caption{adding\newline list(2,4)}\end{subfigure}%
   \begin{subfigure}[b]{20mm}\includegraphics[width=20mm]{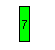}\caption{removing\newline list(0,6) ,\newline list(2,4)}\end{subfigure}%
  \newline\begin{subfigure}[b]{20mm}\includegraphics[width=20mm]{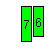}\caption{adding\newline list(2,6)}\end{subfigure}%
  \begin{subfigure}[b]{20mm}\includegraphics[width=20mm]{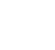}\caption{removing\newline list(2,6),\newline list(1,7)}\end{subfigure}%
  \begin{subfigure}[b]{20mm}\includegraphics[width=20mm]{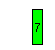}\caption{adding\newline list(2,7)}\end{subfigure}%
  \begin{subfigure}[b]{20mm}\includegraphics[width=20mm]{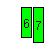}\caption{adding\newline list(1,6)}\end{subfigure}%
  \begin{subfigure}[b]{20mm}\includegraphics[width=20mm]{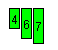}\caption{adding\newline list(0,4)}\end{subfigure}%
  
  \caption{Sorting the sequence $7,\; 6,\; 4$ using the sorting solver introduced in Section \ref{sec:chrintro}.}
  \label{fig:sorting-steps-jawaa}
\end{figure}

\begin{figure}[!ht]
  \centering              
\begin{subfigure}[b]{20mm}\includegraphics[width=20mm]{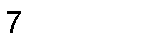}\caption{adding\newline list(0,7)}\end{subfigure}%
  \begin{subfigure}[b]{20mm}\includegraphics[width=20mm]{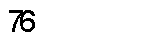}\caption{adding\newline list(1,6)}\end{subfigure}%
  \begin{subfigure}[b]{20mm}\includegraphics[width=20mm]{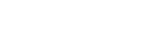}\caption{removing list(0,7) ,\newline list(1,6)}\end{subfigure}%
   \begin{subfigure}[b]{20mm}\includegraphics[width=20mm]{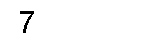}\caption{adding\newline list(1,7)}\end{subfigure}%
   \newline\begin{subfigure}[b]{20mm}\includegraphics[width=20mm]{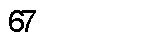}\caption{adding\newline list(0,6)}\end{subfigure}%
  \begin{subfigure}[b]{20mm}\includegraphics[width=20mm]{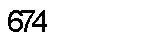}\caption{adding\newline list(2,4)}\end{subfigure}%
   \begin{subfigure}[b]{20mm}\includegraphics[width=20mm]{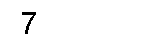}\caption{removing\newline list(0,6) ,\newline list(2,4)}\end{subfigure}%
  \newline\begin{subfigure}[b]{20mm}\includegraphics[width=20mm]{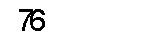}\caption{adding\newline list(2,6)}\end{subfigure}%
  \begin{subfigure}[b]{20mm}\includegraphics[width=20mm]{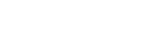}\caption{removing\newline list(2,6),\newline list(1,7)}\end{subfigure}%
  \begin{subfigure}[b]{20mm}\includegraphics[width=20mm]{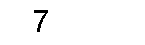}\caption{adding\newline list(2,7)}\end{subfigure}%
  \begin{subfigure}[b]{20mm}\includegraphics[width=20mm]{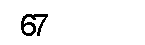}\caption{adding\newline list(1,6)}\end{subfigure}%
  \begin{subfigure}[b]{20mm}\includegraphics[width=20mm]{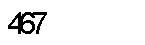}\caption{adding\newline list(0,4)}\end{subfigure}%
   \caption{Sorting the sequence $7,\; 6,\; 4$ using the sorting solver introduced in Section \ref{sec:chrintro}.}
  \label{fig:sorting-steps-jawaa-text}
\end{figure}

For the query \verb+list(7,0), list(6,1),list(4,2)+, the generated animation file after processing the query is shown in \ref{sec:app1}.
As seen from the file, each time a new \verb+list+ constraint was generated, the corresponding node was added. 
The first generated constraint \verb+list(0,7)+ adds to the file:\\\verb-node node7 2 50 10 35 1 7 black green black RECT - which adds a node with x-coordinate: \verb-2-, y-coordinate: \verb-50-, width: \verb-10-, height \verb-35-. The text in the node is \verb-7-.
The execution of the solver keeps on adding \verb+list+ constraints and thus Jawaa nodes are added. In addition, everytime two elements are swapped, their corresponding nodes are removed since the rule \verb+sortlist+ is a simplification rule. The resulting animation is shown in Figure \ref{fig:sorting-steps-jawaa}.
\begin{lstlisting}[language=xml]
<association>
<constraint name="list(Index,Value)">
<add name="text" parameters="name=nodevalueOf(arg1)#x=valueOf(arg0)*12+2#y=50#text=valueOf(arg1)#color=black#size=30" type="Object"/>
</constraint>
</association>
\end{lstlisting}
Figure \ref{fig:sorting-steps-jawaa-text} shows the resulting animation using the query \verb+list(0,7), list(1,6), list(0,4)+.

\section{Conclusion}
\label{sec:conc}
The paper introduced a transformation approach that is able to add visualization features to \chr\ solvers. The new system overcomes the drawbacks of the old approach.
As seen, the need for explicit inputs is removed. The system was able to map constraints into generic existing objects. There was also no need to change the compiler or the runtime system.
We are currently in the process of producing a portable version of the application. It will provide users with running examples in addition to patterns for the annotation module.
Throughout the paper, Jawaa was used for proof of concept. 
However, for the future, the system will also be tested with different tools such as OpenSCAD.
This approach can be used with an existing visualization system since the intelligence is moved to the transformer and not the tracer.
The possibility of using the system to visualize different \chr\ semantics should be examined. The final goal is to have a generic transformation workbench for \textsf{CHR}.
The possibility of having data about the annotations should be also studied. In such a case, the combination of constraints could be used to fire an event.
\bibliography{visualizationforchr}

\begin{thebibliography}{}

\bibitem[\protect\citeauthoryear{Abdennadher and Sharaf}{Abdennadher and
  Sharaf}{2012}]{stschrpaper}
{\sc Abdennadher, S.} {\sc and} {\sc Sharaf, N.} 2012.
\newblock {Visualization of CHR through Source-to-Source Transformation}.
\newblock In {\em ICLP (Technical Communications)}, {A.~Dovier} {and} {V.~S.
  Costa}, Eds. LIPIcs, vol.~17. Schloss Dagstuhl - Leibniz-Zentrum fuer
  Informatik, 109--118.

\bibitem[\protect\citeauthoryear{Duck, Stuckey, Garc\'{\i}a de~la Banda, and
  Holzbaur}{Duck et~al\mbox{.}}{2004}]{refined}
{\sc Duck, G.~J.}, {\sc Stuckey, P.~J.}, {\sc Garc\'{\i}a de~la Banda, M.~J.},
  {\sc and} {\sc Holzbaur, C.} 2004.
\newblock {The Refined Operational Semantics of Constraint Handling Rules}.
\newblock In {\em ICLP}, {B.~Demoen} {and} {V.~Lifschitz}, Eds. Lecture Notes
  in Computer Science, vol. 3132. Springer, 90--104.

\bibitem[\protect\citeauthoryear{Fr{\"u}hwirth}{Fr{\"u}hwirth}{1998}]{Fru98}
{\sc Fr{\"u}hwirth, T.} 1998.
\newblock {Theory and Practice of Constraint Handling Rules, Special Issue on
  Constraint Logic Programming}.
\newblock {\em Journal of Logic Programming\/}~{\em 37,\/}~1-3 (October),
  95--138.

\bibitem[\protect\citeauthoryear{Fr{\"u}hwirth}{Fr{\"u}hwirth}{2008}]{fru_welc%
ome_lnai08}
{\sc Fr{\"u}hwirth, T.} 2008.
\newblock Welcome to {C}onstraint {H}andling {R}ules.
\newblock 1--15.

\bibitem[\protect\citeauthoryear{Fr{\"u}hwirth}{Fr{\"u}hwirth}{2009}]{chrbook}
{\sc Fr{\"u}hwirth, T.} 2009.
\newblock {\em Constraint {H}andling {R}ules}.
\newblock Cambridge University Press.

\bibitem[\protect\citeauthoryear{Fr{\"u}hwirth and Holzbaur}{Fr{\"u}hwirth and
  Holzbaur}{2003}]{stsforexpresive}
{\sc Fr{\"u}hwirth, T.} {\sc and} {\sc Holzbaur, C.} 2003.
\newblock {Source-to-Source Transformation for a Class of Expressive Rules}.
\newblock In {\em APPIA-GULP-PRODE}, {F.~Buccafurri}, Ed. 386--397.

\bibitem[\protect\citeauthoryear{Hundhausen, Douglas, and Stasko}{Hundhausen
  et~al\mbox{.}}{2002}]{DBLP:journals/vlc/HundhausenDS02}
{\sc Hundhausen, C.}, {\sc Douglas, S.}, {\sc and} {\sc Stasko, J.} 2002.
\newblock {A Meta-Study of Algorithm Visualization Effectiveness}.
\newblock {\em Journal of Visual Languages \& Computing\/}~{\em 13,\/}~3,
  259--290.

\bibitem[\protect\citeauthoryear{Kerren and Stasko}{Kerren and
  Stasko}{2002}]{bookintro}
{\sc Kerren, A.} {\sc and} {\sc Stasko, J.} 2002.
\newblock {Chapter 1 Algorithm Animation}.
\newblock In {\em Software Visualization}, {S.~Diehl}, Ed. Lecture Notes in
  Computer Science, vol. 2269. Springer Berlin / Heidelberg, 1--15.

\bibitem[\protect\citeauthoryear{Loveman}{Loveman}{1977}]{Loveman:1977:PIS:321%
992.322000}
{\sc Loveman, D.~B.} 1977.
\newblock {Program Improvement by Source-to-Source Transformation}.
\newblock {\em J. ACM\/}~{\em 24}, 121--145.

\bibitem[\protect\citeauthoryear{Pierson and Rodger}{Pierson and
  Rodger}{1998}]{jawaa}
{\sc Pierson, W.~C.} {\sc and} {\sc Rodger, S.~H.} 1998.
\newblock {Web-based animation of data structures using JAWAA}.
\newblock In {\em Proceedings of the twenty-ninth SIGCSE technical symposium on
  Computer science education}. SIGCSE '98. ACM, New York, NY, USA, 267--271.

\bibitem[\protect\citeauthoryear{Schulte}{Schulte}{1997}]{oz}
{\sc Schulte, C.} 1997.
\newblock {Oz Explorer: A Visual Constraint Programming Tool}.
\newblock In {\em ICLP}, {L.~Naish}, Ed. MIT Press, 286--300.

\bibitem[\protect\citeauthoryear{Simonis and Aggoun}{Simonis and
  Aggoun}{2000}]{tree}
{\sc Simonis, H.} {\sc and} {\sc Aggoun, A.} 2000.
\newblock {Search-Tree Visualisation}.
\newblock In {\em Analysis and Visualization Tools for Constraint Programming},
  {P.~Deransart}, {M.~V. Hermenegildo}, {and} {J.~Maluszynski}, Eds. Lecture
  Notes in Computer Science, vol. 1870. Springer, 191--208.

\end{thebibliography}
\newpage
\appendix
\section{Jawaa Animation File}
\label{sec:app1}
\begin{verbatim}
delay 2500
begin
node node7 2 50 10 35 1 7 black green black RECT  
end
delay 2500
begin
node node6 14 50 10 30 1 6 black green black RECT  
end
delay 2500
begin
remove node7
remove node6
end
delay 2500
begin
node node7 14 50 10 35 1 7 black green black RECT 
end
delay 2500
begin
node node6 2 50 10 30 1 6 black green black RECT 
end
delay 2500
begin
node node4 26 50 10 20 1 4 black green black RECT 
end
delay 2500
begin
remove node6
remove node4
end
delay 2500
begin
node node6 26 50 10 30 1 6 black green black RECT 
end
delay 2500
begin
remove node7
remove node6
end
delay 2500
begin
node node7 26 50 10 35 1 7 black green black RECT 
end
delay 2500
begin
node node6 14 50 10 30 1 6 black green black RECT 
end
delay 2500
begin
node node4 2 50 10 20 1 4 black green black RECT 
end
\end{verbatim}

\section{Jawaa Animation File using Text Annotation Object}
\label{se:app2}
\begin{verbatim}
delay 2500
begin
text node7 2 50 7 black 30 
end
delay 2500
begin
text node6 14 50 6 black 30 
end
delay 2500
begin
remove node7
remove node6
end
begin
text node7 14 50 7 black 30 
end
delay 2500
begin
text node6 2 50 6 black 30 
end
delay 2500
begin
text node4 26 50 4 black 30 
end
delay 2500
begin
remove node6
remove node4
end
begin
text node6 26 50 6 black 30 
end
delay 2500
begin
remove node7
remove node6
end
begin
text node7 26 50 7 black 30 
end
delay 2500
begin
text node6 14 50 6 black 30 
end
delay 2500
begin
remove node4
end
begin
text node4 2 50 4 black 30 
end

\end{verbatim}


\label{lastpage}

\end{document}